\documentclass[aip,amsmath,amssymb,twocolumn,]{revtex4}                

\usepackage{textcase}
\usepackage{graphicx}
\usepackage{dcolumn}
\usepackage{bm}
\usepackage{amsmath,amsthm,amssymb,mathrsfs}

\begin{document}

\title{Indirect excitation of self-oscillation in perpendicular ferromagnet by spin Hall effect}

\author{Tomohiro Taniguchi}

\affiliation{ 
National Institute of Advanced Industrial Science and Technology (AIST), Spintronics Research Center, Tsukuba 305-8568, Japan}

\date{\today}%

\begin{abstract}
A possibility to excite a stable self-oscillation in a perpendicularly magnetized ferromagnet by the spin Hall effect is investigated theoretically. 
It had been shown that such self-oscillation cannot be stabilized solely by the direct spin torque by the spin Hall effect. 
Here, we consider adding another ferromagnet, referred to as pinned layer, on the free layer. 
The pinned layer provides another spin torque through the reflection of the spin current. 
The study shows that the stable self-oscillation is excited by the additional spin torque 
when the magnetization in the pinned layer is tilted from the film plane. 
\end{abstract}

\maketitle


It has been experimentally demonstrated that the spin Hall effect \cite{dyakonov71,hirsch99} in nonmagnetic heavy metals generates pure spin current 
flowing in the direction perpendicular to an external voltage, and excites spin torque on a magnetization in an adjacent ferromagnet 
\cite{kimura07,ando08,seki08,mosendz10,liu12,liu12PRL,pai12,kim13,demidov14,awad17}. 
The spin torque induces the magnetization dynamics such as switching and self-oscillation. 
Substantial efforts have been made to develop practical devices based on the spin Hall effect, 
for example, magnetic random access memory, microwave generator, high sensitivity sensor, 
and new direction such as bio-inspired computing \cite{borders17,kudo17}. 


The spintronics devices based on the spin Hall effect, however, face a serious problem 
because of the geometrical restriction of the spin torque direction. 
Let us assume that an electric current flows in the nonmagnet along $x$ direction, 
while the ferromagnet is set in $z$ direction. 
Then, the direction of the spin polarization in the spin current generated by the spin Hall effect is fixed to $y$ direction. 
The device designs and performances are subject to limitation due to such restriction of spin polarization. 
For example, the magnetization switching of a perpendicular ferromagnet solely by the spin Hall effect is impossible 
because the spin torque does not break the symmetry with respect to the film plane, 
whereas a perpendicular ferromagnet is suitable for a high density memory. 
Using external magnetic field \cite{liu12}, tilted anisotropy \cite{you15,torrejon15}, or exchange bias \cite{fukami16} has been proposed to overcome this issue. 
It was also shown that an excitation of the self-oscillation in a perpendicular ferromagnet solely 
by the spin Hall effect is impossible due to the symmetry \cite{taniguchi15PRB}, 
although a large amplitude oscillation excited in a perpendicular ferromagnet is preferable for an enhancement of emission power. 
Contrary to the case of the switching, this problem has not been solved yet. 


The purpose of this letter is to investigate the possibility to excite the self-oscillation in a perpendicular ferromagnet by the spin Hall effect. 
The work is motivated by recent theoretical studies on the spin-orbit torque in the presence of an additional ferromagnet to the free layer \cite{chen13,taniguchi16IEEE,go17}. 
These theories predict the existence of additional torques and/or enhancement of the spin accumulation. 
Here, we consider adding another ferromagnet, referred to as the pinned layer, on the top of the free layer. 
The pinned layer provides an additional spin torque due to the reflection of the spin current at the interface and the diffusion in bulk. 
This additional torque has a different angular dependence from the conventional spin-orbit torque, 
and results in the excitation of the self-oscillation. 
In the following, we describe the system in this study 
and show the spin torque formula applied in the geometry. 
Then, we investigate the magnetization dynamics by solving the Landau-Lifshitz-Gilbert (LLG) equation numerically. 
It is shown that the self-oscillation can be excited when the magnetization in the pinned layer is tilted from the film plane.



\begin{figure}
\centerline{\includegraphics[width=1.0\columnwidth]{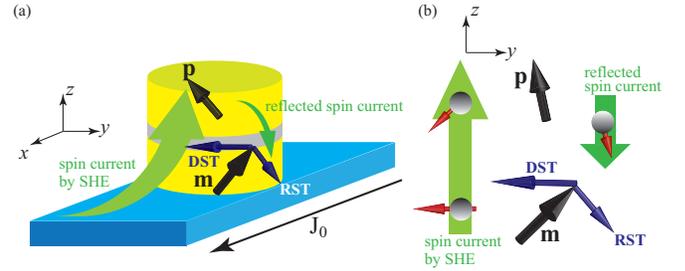}}
\caption{
         (a) Schematic view of the system in this study. 
         The spin Hall effect (SHE) in the bottom nonmagnet injects pure spin current into the free layer. 
         The spin current reflected by the pinned layer is again injected into the free layer 
         The direct and reflection spin torques are referred to as DST and RST, for simplicity. 
         (b) Schematic view of the flow of the spin current and the direction of spin torques. 
         Passing through the free layer from bottom to top, the spin polarization transverse to $\mathbf{m}$ is absorbed and excites direct spin torque. 
         The spin polarization of the reflected spin current is parallel (or antiparallel) to $\mathbf{p}$. 
         The transverse component of the reflected spin current is absorbed to the free layer and exictes the reflection spin torque.
         \vspace{-3ex}}
\label{fig:fig1}
\end{figure}




The system under consideration is schematically shown in Fig. \ref{fig:fig1}(a). 
The bottom layer is a nonmagnetic heavy metal showing the spin Hall effect. 
Applying an external voltage along the in-plane ($x$) direction, 
the electric current with the density $J_{0}$ is converted to pure spin current flowing in $z$ direction. 
We also consider placing a pinned layer onto the free layer. 
The spin current excites the magnetization dynamics in the free layer through the spin-transfer effect. 
We denote the unit vectors pointing in the magnetization direction of the free and pinned layers as $\mathbf{m}$ and $\mathbf{p}$, respectively. 
The magnetization dynamics in the free layer is described by the LLG equation as 
\begin{equation}
  \frac{d \mathbf{m}}{dt}
  =
  -\gamma
  \mathbf{m}
  \times
  \mathbf{H}
  +
  \mathbf{T}
  +
  \alpha
  \mathbf{m}
  \times
  \frac{d \mathbf{m}}{dt},
  \label{eq:LLG}
\end{equation}
where $\gamma$ and $\alpha$ are the gyromagnetic ratio and the Gilbert damping constant, respectively. 
The magnetic field $\mathbf{H}$ consists of a perpendicular anisotropy field 
and the stray field from the pinned layer, 
and is given by 
\begin{equation}
  \mathbf{H}
  =
  -H_{\rm d}
  p_{y}
  \mathbf{e}_{y}
  +
  \left[
    2 H_{\rm d}
    p_{z}
    +
    \left(
      H_{\rm K}
      -
      4 \pi M 
    \right)
    m_{z}
  \right]
  \mathbf{e}_{z},
  \label{eq:field}
\end{equation}
where $H_{\rm d}$ characterizes the magnitude of the stray field, 
whereas $H_{\rm K}$ and $4\pi M$ with the saturation magnetization $M$ are the crystalline and shape (demagnetization) fields, respectively. 
The magnitude of the stray field found in the experiments is typically on the order 100 Oe \cite{liu12}, 
which is consistent with a theoretical evaluation; see Supplementary Material. 
Thus, in this paper, we use the value of $H_{\rm d}=100$ Oe in the following calculations. 
We assume that $H_{\rm K}>4\pi M$, and 
the free layer consequently becomes perpendicularly magnetized in the absence of the pinned layer. 
The spin torque in Eq. (\ref{eq:LLG}) is $\mathbf{T}$, 
which in the present geometry is given by 
\begin{equation}
\begin{split}
  \mathbf{T}
  =&
  -\frac{\hbar \eta_{1} J_{0}}{2e Md}
  \mathbf{m}
  \times
  \left(
    \mathbf{e}_{y}
    \times
    \mathbf{m}
  \right)
\\
  &
  -
  \frac{\hbar \eta_{2} J_{0}}{2eMd}
  \frac{m_{y} \left( \mathbf{m}\cdot\mathbf{p} \right) [\mathbf{m} \times \left( \mathbf{p} \times \mathbf{m} \right)]}{1 - \lambda^{2}(\mathbf{m}\cdot\mathbf{p})^{2}}, 
  \label{eq:spin_torque_def}
\end{split}
\end{equation}
where $d$ is the thickness of the free layer, 
whereas $e(>0)$ is the elementary charge. 
The first term on the right hand side of Eq. (\ref{eq:spin_torque_def}) is the conventional spin Hall torque 
directly excited by the spin current generated by the bottom nonmagnet. 
For convention, we call this torque the direct spin torque; see Fig. \ref{fig:fig1}(a). 
On the other hand, the second term arises from the spin current transmitted through the free layer 
and reflected by the pinned layer, 
which is also schematically shown in Fig. \ref{fig:fig1}(a). 
In a same manner, we call this torque the reflection spin torque.  


Before solving the LLG equation, let us explain the physical meaning, as well as its derivation, of these spin torques in this system. 
The spin torque has been calculated theoretically 
by using several methods such as the ballistic spin transport theory with the interface scattering \cite{slonczewski96,brataas01}, 
the first-principles calculations \cite{stiles02,zwierzycki05}, 
the Boltzmann approach \cite{stiles02JAP,xiao04}, and the diffusive spin transport theory in bulk \cite{zhang02,taniguchi09}. 
Although the parameters characterizing the spin torque depend on the models, 
these theories basically deduce the same angular dependence of the spin torque. 
The derivation of Eq. (\ref{eq:spin_torque_def}) in the present geometry 
using the diffusive spin transport theory in bulk and the interface scattering theory is summarized in Supplementary Material. 
The spin torque efficiency $\eta_{1}$ of the direct spin torque is proportional to the spin Hall angle in the bottom nonmagnet. 
It also depends on the interface and bulk properties. 
Note that the direct spin torque is excited by an absorption of the transverse component (perpendicular to $\mathbf{m}$) 
of a spin current generated by the spin Hall effect in the bottom nonmagnet. 
The spin polarization of the spin current points to the $y$ direction, and thus, 
the direct spin torque moves the magnetization parallel or antiparallel to the $y$ axis, as schematically shown in Fig. \ref{fig:fig1}(b). 


On the other hand, the reflection spin torque arises from the spin current passing through the free layer. 
Such spin current is again injected into the free layer from the top interface due to the reflection from the top interface 
and diffusive spin transport in the pinned layer. 
Note that the spin current generated in the bottom nonmagnet has the spin polarization in the $y$ direction. 
Because of the absorpition of the transverse component of the spin current from the bottom nonmagnet mentioned above, 
the reflection spin torque includes the factor $m_{y}$; 
i.e., when $m_{y}=0$, the spin current generated in the bottom nonmagnet is completely absorbed to the free layer at the bottom interface, 
and therefore, the reflection spin torque becomes zero because the spin current passing through the free layer is unpolarized. 
Similarly, the spin polarization parallel to $\mathbf{p}$ survives during the transport through the pinned layer. 
As a result, the reflection spin torque also includes the factor $\mathbf{m}\cdot\mathbf{p}$ on the numerator in Eq. (\ref{eq:spin_torque_def}). 
Moreover, the direction of the reflection spin torque is given by $\mathbf{m}\times(\mathbf{p}\times\mathbf{m})$, as schematially shown in Fig. \ref{fig:fig1}(b), 
in comparison to that of the direct spin torque pointing to the direction of $\mathbf{m}\times(\mathbf{e}_{y}\times\mathbf{m})$. 
The spin torque efficiency $\eta_{2}$ and the parameter $\lambda$ determining the angular dependence 
characterize the amount of the spin current reinjected from the top interface to the free layer. 
Their values depend not only on the spin Hall angle in the bottom nonmagnet 
but also on the interface and bulk properties of the pinned layer, 
such as spin diffusion length and mixing conductance. 
The details of the derivation of the reflection spin torque, as well as the relation to material parameters in the diffusive model, 
are summarized in Supplementary Material.


We note that the present model is applicable to a metallic multilayer. 
When a spacer between the free and pinned layers is replaced by an oxide barrier, as in the case of a magnetic tunnel junction, 
the spin current cannot penetrate into the pinned layer, and thus, the reflection spin torque becomes zero. 
When an electric voltage is applied along the perpendicular direction, 
as in the case of the experiment to obtain an electric signal through tunnel magnetoresistance effect \cite{liu12}, 
a spin current will be driven between the free and pinned layer, 
and a torque similar to the reflection spin torque will appear. 
The angular dependence of such a torque, however, might be different from the reflection spin torque.


We investigate the magnetization dynamics in this geometry by solving Eq. (\ref{eq:LLG}) numerically. 
The values of the parameters are brought from typical experimental values in spin torque oscillator \cite{kubota13}, 
i.e., $M=1448.3$ emu/c.c., $H_{\rm K}=18.6$ kOe, $d=2$ nm, $\gamma=1.764 \times 10^{7}$ rad/(Oe s), and $\alpha=0.005$. 
The spin torque parameters are $\eta_{1}=0.14$, $\eta_{2}=0.07$, and $\lambda=0.82$, respectively; see Supplementary Material for the evaluations of these parameters. 
The magnetization in the pinned layer is 
\begin{equation}
  \mathbf{p}
  =
  \begin{pmatrix}
    0 \\
    -\sin\theta_{\rm p} \\
    \cos\theta_{\rm p}
  \end{pmatrix}, 
\end{equation}
where $\theta_{\rm p}$ is the tilted angle from $z$ axis. 
We note that efforts have been made to realize a tilted state ($\theta_{\rm p} \neq 0^{\circ}$ nor $90^{\circ}$) of a magnetization in a ferromagnet 
by making use of a higher-order anisotropy or an interlayer exchange coupling between a perpendicular and an in-plane magnetized ferromagnets \cite{lee02,zha09,stillrich09,fallarino16}. 
The initial state is the energetically stable state given by $\mathbf{m}(0)=(0,\sin\theta_{0},\cos\theta_{0})$, 
where $\theta_{0}$ is the tilted angle of the magnetization from $z$ axis, which minimizes the energy density given by 
$E=-M \int d \mathbf{m}\cdot\mathbf{H}=-M [H_{\rm d} \sin\theta_{\rm p} m_{y} + 2 H_{\rm d} \cos\theta_{\rm p} m_{z} + (H_{\rm K}-4\pi M) m_{z}^{2}/2]$. 
We note that the magnetization in equilibrium is destabilized by the spin torques 
when the current density is larger than a critical value given by 
\begin{equation}
  J_{\rm c}
  =
  \frac{2 \alpha eMd}{\hbar \mathcal{P}}
  \left(
    \frac{H_{X} + H_{Y}}{2}
  \right), 
  \label{eq:critical_current}
\end{equation}
where an effective spin polarization $\mathcal{P}$ is derived as 
\begin{equation}
  \mathcal{P}
  =
  \eta_{1}
  \sin\theta_{0}
  +
  \frac{\eta_{2}}{1-\lambda^{2} p_{Z}^{2}}
  \left(
    p_{Z}^{2}
    \sin^{2}\theta_{0}
    -
    \frac{\xi}{2}
    p_{X}
  \right).
\end{equation}
Here, $p_{Z}=-\sin(\theta_{\rm p}+\theta_{0})$ and $p_{X}=\cos(\theta_{\rm p}+\theta_{0})$, 
whereas $\xi=\Lambda p_{Z} \sin\theta_{0} + p_{X} \sin\theta_{0} + p_{Z} \cos\theta_{0}$ with 
$\Lambda= 2\lambda^{2} p_{X}p_{Z}/(1-\lambda^{2}p_{Z}^{2})$. 
The fields $H_{X}$ and $H_{Y}$ in Eq. (\ref{eq:critical_current}) are expressed as 
\begin{equation}
\begin{split}
  H_{X}
  =&
  H_{\rm d}
  \left(
    2 \cos\theta_{\rm p}
    \cos\theta_{0}
    +
    \sin\theta_{\rm p}
    \sin\theta_{0}
  \right)
\\
  &+
  \left(
    H_{\rm K}
    -
    4\pi M
  \right)
  \cos 2 \theta_{0},
\end{split}
\end{equation}
\begin{equation}
\begin{split}
  H_{Y}
  =&
  H_{\rm d}
  \left(
    2 \cos\theta_{\rm p}
    \cos\theta_{0}
    +
    \sin\theta_{\rm p}
    \sin\theta_{0}
  \right)
\\
  &+
  \left(
    H_{\rm K}
    -
    4\pi M
  \right)
  \cos^{2} \theta_{0}.
\end{split}
\end{equation}
We note that the ferromagnetic resonance (FMR) frequency is related to $H_{X}$ and $H_{Y}$ as 
\begin{equation}
  f_{\rm FMR}
  =
  \frac{\gamma}{2\pi}
  \sqrt{
    H_{X}
    H_{Y}
  }.
  \label{eq:FMR}
\end{equation}
The derivation of Eq. (\ref{eq:critical_current}) based on the linearized LLG equation is summarized in Supplementary Material. 
Equation (\ref{eq:critical_current}) diverges when the magnetization in the pinned layer points to the perpendicular direction, $\theta_{\rm p}=0$. 
This fact indicates that the linearized LLG equation is inapplicable to study the instability analysis of the magnetization dynamics. 
In this case, the critical current will be independent of the damping constant, as studied in Ref. \cite{lee13}, 
and does not show self-oscillation.



\begin{figure}
\centerline{\includegraphics[width=1.0\columnwidth]{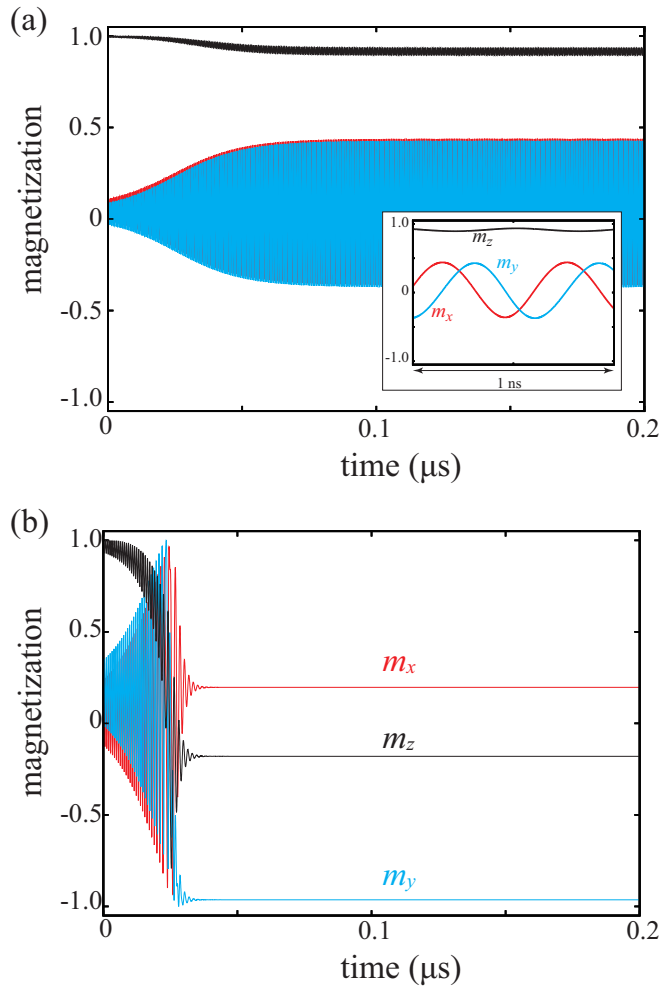}}
\caption{
         Time evolutions of the magnetization ($m_{x}$ in red, $m_{y}$ in blue, and $m_{z}$ in black) for (a) $\theta_{\rm p}=10^{\circ}$ and (b) $60^{\circ}$. 
         The current density is $20$ MA/cm${}^{2}$. 
         The inset in (a) shows the oscillation of the magnetization in a steady state. 
         \vspace{-3ex}}
\label{fig:fig2}
\end{figure}




Figures \ref{fig:fig2}(a) and \ref{fig:fig2}(b) show examples of the magnetization dynamics obtained from Eq. (\ref{eq:LLG}), 
where $\theta_{\rm p}$ is $10^{\circ}$ in (a) and $60^{\circ}$ in (b). 
The current density is $20$ MA/cm${}^{2}$ in these calculations, 
whereas the critical current density estimated from Eq. (\ref{eq:critical_current}) is $9.5$ MA/cm${}^{2}$ for $\theta_{\rm p}=10^{\circ}$ 
and $7.4$ MA/cm${}^{2}$ for $\theta_{\rm p}=60^{\circ}$. 
As shown, a stable oscillation is excited for $\theta_{\rm p}=10^{\circ}$; 
this is the main finding in this study. 
The magnetization precesses around an axis slightly tilted from $z$ axis. 
The oscillation frequency is 1.60 GHz, which is slightly smaller than the FMR frequency, 1.72 GHz. 
The relaxation time to the self-oscillation time is about 50 ns. 
We note here that the inverse of the relaxation time is proportional to the current \cite{taniguchi17}, 
and therefore, the relaxation time will be shortened by applying a large current. 
On the other hand, when $\theta_{\rm p}=60^{\circ}$, 
the magnetization switches to the direction antiparallel to the $y$ direction without showing a self-oscillation, 
as shown in Fig. \ref{fig:fig2}(b), 
for a current larger than the critical current. 
This behavior is similar to that excited solely by the direct spin torque studied in Ref. \cite{taniguchi15PRB}.



\begin{figure}
\centerline{\includegraphics[width=1.0\columnwidth]{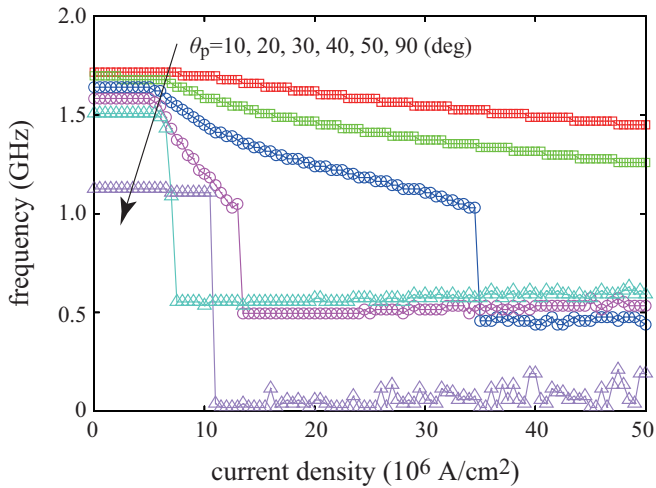}}
\caption{
         Current dependences of the oscillation frequency of the magnetization at finite temperature for 
         $\theta_{\rm p}=10^{\circ}$ (red square), $20^{\circ}$ (green square), $30^{\circ}$ (blue circle), $40^{\circ}$ (magenta circle), 
         $50^{\circ}$ (turquoise triangle), and $90^{\circ}$ (purple triangle). 
         The critical current densities at zero temperature for these $\theta_{\rm p}$ are $9.5$, $6.3$, $5.8$, $6.1$, $6.7$, and $10.1$ MA/cm${}^{2}$. 
         \vspace{-3ex}}
\label{fig:fig3}
\end{figure}



The current dependences of the oscillation frequency for several values of $\theta_{\rm p}$ are summarized in Fig. \ref{fig:fig3}. 
Random torque, $-\gamma\mathbf{m}\times\mathbf{h}$, originated from thermal fluctuation is added to the right hand side of Eq. (\ref{eq:LLG}) 
to evaluate the magnoise frequency below the threshold. 
The components of the random torque satisfy the fluctuation-dissipation theorem, 
$\langle h_{k}(t) h_{\ell}(t^{\prime}) \rangle = [2 \alpha k_{\rm B}T/(\gamma MV)] \delta_{k \ell} \delta(t-t^{\prime})$, 
where the temperature $T$ and the cross-section area $S$ ($V=Sd$) are assumed as 300 K and $\pi \times 60^{2}$ nm${}^{2}$ \cite{kubota13}, respectively. 
The oscillation frequency is estimated from the Fourier transformation of $m_{y}(t)$, where the spectra are averaged over $10^{3}$ realizations. 
When the current density is smaller than the critical current density, 
the magnetization oscillates around the equilibrium state, and thus, the magnoise appears around the FMR frequency. 
When $\theta_{\rm p}=90^{\circ}$ and the current is larger than the critical value, 
the magnetization switches its direction to the negative $y$ direction without showing a self-oscillation, 
similar to that shown in Fig. \ref{fig:fig2}(b). 
As a result, a discontinuous change of the oscillation frequency appears near the critical value, $J_{\rm c}=10.1$ MA/cm${}^{2}$. 
For $\theta_{\rm p}=30^{\circ}$, $40^{\circ}$, and $50^{\circ}$, 
the magnetization shows the self-oscillation when the current is larger than the critical value. 
In the self-oscillation, the oscillation frequency decreases with increasing current magnitude. 
Above certain values of the current, however, the magnetization switching occurs, 
and thus, the discontinuous drops of the oscillation frequency are observed. 
On the other hand, self-oscillations are observed for the present range of the current ($J_{0} \le 50$ MA/cm${}^{2}$) 
when $\theta_{\rm p}=10^{\circ}$ and $20^{\circ}$ 
(a switching for $\theta_{\rm p}=10^{\circ}$ occurs at a sufficiently large current $J_{0}>115$ MA/cm${}^{2}$). 
These results indicate that the self-oscillation is stably excited 
when the magnetization in the pinned layer is tilted, particularly in close range, from the perpendicular ($z$) axis. 
A possible reason why a current over which the stable self-oscillation terminates becomes smaller when $\theta_{\rm p}$ becomes larger is 
due to the characteristics of angular dependence of the reflection spin torque. 
As mentioned above, the angular dependence of the reflection spin torque includes the term $\mathbf{m}\cdot\mathbf{p}$. 
As can be seen in Fig. \ref{fig:fig2}(a), the self-oscillation is excited closely around the $z$ axis. 
Consequently, the magnitude of the reflection spin torque becomes small for a large $\theta_{\rm p}$ ($\mathbf{p} \to \mathbf{e}_{y}$), 
making the effect of the reflection spin torque on the oscillation small and region of the stable self-oscillation narrow. 
We note, however, that the self-oscillation cannot be excited 
when the magnetization in the pinned layer completely points to the $z$ direction, as mentioned above. 


Finally, let us discuss the role of the reflection spin torque in the above results. 
We emphasize that the reflection spin torque plays a key role to stabilize the self-oscillation. 
To understand this argument, we revisit the theoretical conditions to excite the self-oscillation studied in our previous work \cite{taniguchi15PRB}. 


First of all, the spin torque should supply a finite positive energy to the free layer during the oscillation 
to cancel the energy dissipation due to the damping torque. 
In the conventional geometry of the spin Hall devices consisting of a single perpendicular ferromagnet and in the absence of an external field, 
the energy supplied by the spin torque becomes totally zero due to the axial symmetry of the oscillation orbit. 
Therefore, a self-oscillation cannot be excited \cite{taniguchi15PRB}. 


A way to solve this problem is to apply an external magnetic field. 
The field breaks the symmetry of the oscillation orbit, 
and makes the supplied energy by the spin torque finite. 
In this work, the stray field from the pinned layer plays the role of such external field. 
This is, however, not sufficient enough to stabilize the self-oscillation. 
The second condition necessary to stabilize the self-oscillation is that 
a current magnitude should be larger than the critical current destabilizing the equilibrium state \cite{taniguchi15PRB}. 
If this condition is unfulfilled, the free layer undergoes the magnetization switching above the critical current 
without showing a self-oscillation \cite{taniguchi15PRB}, 
as in the case shown in Fig. \ref{fig:fig2}(b). 


It was shown in Ref. \cite{taniguchi15PRB} that the direct spin torque is not sufficient to stabilize the self-oscillation in the spin Hall geometry 
because the second condition is not satisfied even in the presence of an external field. 
On the other hand, in the present study, the self-oscillation is excited, as shown in Fig. \ref{fig:fig2}(a). 
This fact indicates that the reflection spin torque fulfills the second condition, 
and stabilizes the self-oscillation. 



One might be interested in proving the stabilization of the self-oscillation by the reflection spin torque analytically, 
instead of the numerical approach done in the above calculations. 
It is relatively easy to confirm whether the first condition to stabilize the self-oscillation is satisfied 
by focusing on the symmetries of the oscillation orbit and the angular dependence of the spin torque. 
On the other hand, as far as we know, it cannot be easily confirmed to fulfill the second condition. 
It should individually be examined for each system. 
In principle, the second condition can be studied theoretically by deriving an analytical formula of $\mathbf{m}$ corresponding to an oscillation orbit, 
and solving the energy balance equation \cite{bertotti09}. 
Then, it becomes, for example, possible to derive analytical conditions on the material parameters to stabilize the self-oscillation. 
These calculations are, however, generally complicated in practice, except for a few cases. 
In the present system, the solution of the oscillation orbit can be described by the elliptic functions, in principle. 
It involves, however, complex mathematics, and thus, 
analytical calculations to study the satisfaction of the second condition are beyond the scope of this paper. 





In conclusion, the magnetization dynamics in the spin Hall geometry in the presence of an additional ferromagnet was studied theoretically. 
In addition to the direct spin torque by the spin Hall effect, 
the additional ferromagnet provides another spin torque through the reflection of the spin current. 
Solving the LLG equation with the direct and reflection spin torques numerically, 
it was found that a stable self-oscillation can be excited when the magnetization in the pinned layer is tilted from the film-plane. 


\section*{Supplementary Material}

Supplementary Material includes the derivations of Eqs. (\ref{eq:field}), (\ref{eq:spin_torque_def}), and (\ref{eq:critical_current}). 

\section*{Acknowledgement}

The author is grateful to Takehiko Yorozu and Hitoshi Kubota for valuable discussion. 
The author is also thankful to Satoshi Iba, Aurelie Spiesser, Hiroki Maehara, and Ai Emura 
for their support and encouragement. 
This work was supported by JSPS KAKENHI Grant-in-Aid for Young Scientists (B) 16K17486. 








\end{document}